# InP-based single-photon sources operating at telecom C-band with increased extraction efficiency

A. Musiał[1,a)], M. Mikulicz[1], P. Mrowiński[1], A. Zielińska[1], P. Sitarek[1], P. Wyborski[1], M. Kuniej[1], J. P. Reithmaier[2], G. Sęk[1], and M. Benyoucef[2,a)]

[1]*Department of Experimental Physics, Faculty of Fundamental Problems of Technology, Wrocław University of Science and Technology, Wybrzeże Wyspiańskiego 27, 50-370 Wrocław, Poland*

[2]*Institute of Nanostructure Technologies and Analytics (INA), Center for Interdisciplinary Nanostructure Science and Technology (CINSaT), University of Kassel, Heinrich-Plett-Str. 40, 34132 Kassel, Germany*

**Abstract**

In this work we demonstrate a triggered single-photon source operating at the telecom C-band with photon extraction efficiency exceeding any reported values in this range. The non-classical light emission with low probability of the multiphoton events is realized with single InAs quantum dots (QDs) grown by molecular beam epitaxy and embedded directly in an InP matrix. Low QD spatial density on the order of $5\times10^8$ cm$^{-2}$ to $\sim2\times10^9$ cm$^{-2}$ and symmetric shape of these nanostructures together with spectral range of emission makes them relevant for quantum communication applications. The engineering of extraction efficiency is realized by combining a bottom distributed Bragg reflector consisting of 25 pairs of InP/In$_{0.53}$Ga$_{0.37}$Al$_{0.1}$As layers and cylindrical photonic confinement structures. Realization of such technologically non-demanding approach even in a non-deterministic fashion results in photon extraction efficiency of $(13.3\pm2)\%$ into 0.4 numerical aperture detection optics at approx. 1560 nm emission wavelength, i.e., close to the center of the telecom C-band.

---

a) Corresponding authors: anna.musial@pwr.edu.pl, m.benyoucef@physik.uni-kassel.de

Epitaxial nanostructures are the most promising realization of single-photon sources (SPSs) in terms of single-photon purity [Musiał et al. (2020), Arakawa et al. (2020), Schweickert et al. (2018), Michler ed. (2017), Miyazawa et al. (2016)]. The probability of multiphoton emission events as low as $(7.5\pm1.6)\times10^{-5}$ and $(4.4\pm0.2)\times10^{-4}$ was achieved at visible and telecom wavelengths, respectively. These are values unreachable by any other solution and crucial in view of most of the quantum information and communication technologies [Arakawa et al. (2020), Michler ed. (2017), Lodahl et al. (2018), Gisin et al. (2006), Knill et al. (2001)]. Quantum dot (QD) based SPS with emission wavelength below 1 μm have reached almost ideal characteristics [Huber et al. (2018), Somaschi et al. (2016), Ding et al. (2016), Unsleber et al. (2016), Schlehahn et al. (2015), Gschrey et al. (2015), Gazzano et al. (2013), Claudon et al. (2010), Heindel et al. (2010)] and there is no fundamental limitation for their long-wavelength counterparts to follow the same path, but further developments and optimizations are required to fulfill this goal. The main advantage of the latter is the operating spectral range which allows for long distance data transmission in the fiber [Cao et al. (2019)] due to lack of distortion (O-band at 1310 nm) and minimal attenuation of optical signals (C-band at 1550 nm).

One of the issues that needs to be overcome in the case of epitaxial nanostructures is the extraction of photons from the semiconductor matrix which is limited by the total internal reflection on the air-semiconductor interface at the sample surface and distribution of the electromagnetic radiation in the far field (omnidirectional emission pattern). In the case of a QD in a homogeneous medium only approx. 1% of all emitted photons can be collected by the first lens of the detection system (following definition of extraction efficiency) [Semenova et al. (2008), Barnes et al. (2002)] which limits the generation rate of QD-based SPSs (and therefore, e.g., bit rate of the long distance transmission) as well as enforces higher excitation in the quantum optical experiments degrading the coherence properties of emitted photons [Anderson et al. (2020)]. This remains a challenge for the telecom range, in particular for the C-band, on which we will focus from now on. The experimentally obtained extraction efficiency value reported so far for this spectral range is 10.9 % into the numerical aperture of 0.55 for a horn structure [Takemoto et al. (2007)] in InAs/InP material system. These structures were further used to demonstrate quantum key distribution over 120 km and achieved emission efficiency of 5.8% (corrected for non-ideal $g^{(2)}(0)$ value) [Takemoto et al. (2010)] and emission rate of single photon pulses exceeding 3 MHz [Takemoto et al. (2015)]. This needs to be further optimized for the sources to become truly practical.

The distributed Bragg reflector (DBR) itself is more challenging that for the GaAs-based material system both on the design and fabrication level. This is because materials lattice-matched to InP do not provide as high refractive index contrast as in the case of GaAs/AlAs explored for shorter wavelengths. This can be overcome by combination of complex quaternary compounds and higher number of DBR pairs, which constitutes a technological challenge as their composition and thickness needs to be precisely controlled. Even tough, DBRs with reflectivity as high as 95% have already been realized [T. Müller et al. (2018), Benyoucef et al. (2013)] and Si/SiO$_2$ micropillar cavities with 60% output efficiency theoretically proposed [Song et al., (2013)]. Applying microlenses for increasing the extraction efficiency [Gschrey et al. (2015), Sartison et al. (2018)] has the disadvantage of the size of the photonic structure scaling with the emission wavelength which would lead to mesas exceeding 3.5 μm in diameter [Schneider et al. (2018)] in the case of telecom wavelengths. This would further require very low spatial QD density for a single QD occupation of a microlens, because otherwise presence of other QDs in the same microlens would degrade the output of the source. It is also not very efficient material usage even in the case of deterministic platforms and to overcome this issue spectrally selective excitation of a target QD have to be applied.

In this work we present realization of an approach combining bottom DBR section with a cylindrical mesa for photonic confinement to improve the QD directionality. This allowed exceeding the up-to-date experimentally demonstrated extraction efficiency for the epitaxial nanostructures based SPSs at the telecom C-band. Extraction efficiency of 25% (30%) for numerical aperture of 0.4 (0.55) was theoretically predicted for mesa structure on DBR [Mrowiński et al. (2019)]. Additionally, it is far simpler in fabrication than sophisticated approaches as the horn structures and simultaneously provides broadband enhancement of the extraction efficiency (theoretical full width of half maximum exceeding 60 nm [Mrowiński et al. (2019)]), and therefore does not require exact spectral matching of the QD emission line with the photonic structure spectral characteristics in contrast to still more common approaches utilizing microcavities. We analyze in detail emission of a quantum dot in a cylindrical mesa and demonstrate that both, good single-photon emission properties and increased extraction efficiency can be achieved in such a technologically non-demanding system in the third telecom window.

The structure investigated here was grown by molecular beam epitaxy (MBE) and consists of a single layer of InAs QDs embedded directly in InP matrix and deposited on a (001) InP Fe-doped substrate. The thickness of the surrounding InP epitaxial material is 246 nm. The bottom DBR section underneath is composed of 25 pairs of InP/In$_{0.53}$Ga$_{0.37}$Al$_{0.1}$As layers with nominal thicknesses of 123 nm and 110 nm, respectively. The measured reflectivity spectrum (not shown here) shows 93 nm wide stopband with 1551.5 nm central wavelength at room temperature which translates into 1523.5 nm at low temperature (10K). Additional ripening step applied during the MBE growth of the dots [Yacob et al. (2014), Kamins et al. (1999)] results in their uncommon morphology, i.e. symmetric and relatively large nanostructures (on average 55 nm diameter and up to 15 nm height) of low QD spatial density ($5\times10^8$ cm$^{-2}$ to ~$2\times10^9$ cm$^{-2}$) [Yacob et al. (2014)].

The mesas were realized non-deterministically by means of electron-beam lithography and wet chemical etching. The nominal etching depth was 600 nm and mesas' diameters from 200 nm to 900 nm with 50 nm step were patterned in order to find the best case of a mesa with a single QD close to its center to maximize the extraction efficiency, taking into account low QD spatial density, fabrication accuracies and imperfections.

The optical measurements were realized in a microphotoluminescence (PL) setup with 100 µeV and 2 µm spectral and spatial resolutions, respectively, equipped with 320 mm focal length monochromator, InGaAs CCD and 0.4 numerical aperture long working distance (20 mm) microscope objective. For the extraction efficiency measurement, the setup was extended with the fiber-coupling to a single-photon superconducting nanowire (SNSPD) NbN detector with quantum efficiency (QE) exceeding 90% at 1550 nm. The setup was calibrated using cw semiconductor tunable IR laser which allowed determining the setup efficiency at 1560 nm of 0.8% and 1.2% for vertical and horizontal linear polarizations, respectively. For the photon statistics measurements the setup was further extended: for autocorrelation, a fiber 50:50 beam splitter (BS) and for the cross-correlation a second Hanbury Brown and Twiss interferometer arm was realized with 50:50 non-polarizing cube BS and fiber coupling to a 0.4 Gaussian bandwidth tunable bandpass filter followed by SNSPD with QE over 80%. For all measurements, the sample was cooled down to 4.5K in continuous flow liquid helium cryostat and excited non-resonantly. The optical excitation was provided by a semiconductor laser with 640 nm and 805 nm wavelength for cw and pulsed (80 MHz repetition rate of less than 50 ps long pulses) excitation, respectively.

To experimentally determine the photon extraction efficiency the single QD emission rate under pulsed excitation at saturation power needs to be measured and corrected for the setup efficiency. In Fig. 1a) a series of spectra for various excitation powers under pulsed excitation measured on an SNSPD detector is presented. Two emission lines separated by 1 meV dominate the spectrum. The analysis of the emission rate as a function of excitation power (Fig. 1b)) shows maximal (saturation) emission rate of 50.9 kHz and 58.5 kHz at the detector for lines A and B, respectively (0.95 kHz was demonstrated for Purcell-enhanced single photon source [Birowosuto et al., (2012)]). This corresponds to generation rate of the SPS on the order of 5 MHz.

To evaluate the extraction efficiency the polarization properties of emission need to be determined due to polarization-dependence of the setup transmission. The PL spectra as a function of the linear polarization direction are presented in Fig. 2a) and the polar plot of the intensity for the lines of interest in Fig. 2b). The optical axes for the two emission lines are approximately anti- and diagonal. The degree of the linear polarization equals 31% and 46% for line A and B, respectively, which is consistent with the substantial light hole-heavy hole mixing expected for large height of the investigated nanostructures [Musiał et al. (2012)], even in the case of only slight inherent asymmetry of the confining potential. In the case of photon extraction efficiency from a single QD, the total number of emitted and detected photons are of interest. This means that the number of emitted photons from the mutually excluding excitonic complexes from the same QD should be added up (because part of the excitation pulses results in emission of one complex and the other part – of a different complex), but not the ones within the same cascade (this scenario leads to two photons emitted per excitation pulse).

Excitation power dependence of PL under cw excitation (Fig. 2c) and d)) suggests that the two emission lines does not belong to the same recombination cascade – both exhibit almost linear intensity dependence (slope 1.1 and 1.2 for line A and B, respectively) and are visible in the spectrum even at the lowest excitation. This means the probability of the certain excitonic complexes formation cannot be substantially different and therefore most probably it requires that similarly low number of carriers are captured in the QD. Additionally, the evolution of the emission spectra as a function of linear polarization angle is similar for the two lines and does not show any clear splitting within the given spectral resolution whereas for a cascade (both neutral and charged excitons) an anticorrelated behavior is expected [Kettler et al. (2016)]. Also, the radiative lifetime is similar for both lines and equals 1.2 and 1.3 ns for line A and B, respectively. In the case of biexciton-exciton cascade typically exciton lifetime is at least twice longer in the strong confinement regime [Narvaez et al. (2006)]. For the lifetimes to be equal we should be in the weak confinement regime [Wimmer et al. (2006)] which is rather not the case here as the exciton Bohr radius for bulk InAs (InP) equals to 34 (15) nm and magnetooptical study performed on these structures revealed in-plane extent of the wave function of (12-13) nm [Rudno-Rudziński et al. (2021)].

To prove that the two emission lines indeed originate from excitonic complexes confined in the same QD, the cross-correlation measurements were performed exhibiting clear antibunching confirming that the two investigated decay channels are temporally anti-correlated (Fig. 3a)) and therefore result from recombination from the same QD. The observed bunching, for the negative time delays, indicates that there is a non-zero probability of capturing a single carrier by the QD before exciting the charged exciton again [Reimer et al. (2016), Suffczyński et al. (2006)]. This bunching together with strong asymmetry of the histogram prove that capturing single carriers by the QD is very important in the QD excitation process. This is reflected in high probability of trion formation – previous study of these structures revealed that the highest intensity emission lines originate from recombination of charged excitons [Rudno-Rudziński et al. (2021)]. Taking this into account, the extraction efficiency from the target QD can be evaluated from the sum of emission rates for lines A and B (Fig. 1b)) to be equal to 13.3%. This is the maximal value for all QDs investigated in the sample which means that in that case the QD was indeed located close to the center of the mesa. The actual size of this mesa is 900 nm (measured by atomic force microscopy). The experimentally obtained value lower than theoretical predictions [Mrowiński et al. (2019)] can be a result of a non-ideally central position of the QD within the mesa and the tolerances of the fabrication process. The method of experimental determination of the extraction efficiency assumes 100% efficient excitation as well as ideal internal QE and therefore constitutes the lower limit for the extraction efficiency. Comparison with the results obtained on the planar sample without patterning (6.8% of extraction efficiency obtained experimentally in the best case) brings two conclusions: the internal QE of investigated nanostructures is rather high and the bottom DBR section and mesa are equally responsible for boosting the extraction efficiency for the current sample. This is consistent with one order of magnitude increase in the emission intensity between the sample with and without DBR [Kors et al. (2018)]. Embedding QDs in photonic mesa structure provides further increase of the extraction efficiency, but not as high as expected theoretically. These means that this effect is partially overshadowed by the fabrication imperfections and accuracies, and possibly also due to decrease in QD internal QE caused by close proximity of etched surfaces - rough mesa sidewalls containing carrier traps.

Single-photon emission is proven by autocorrelation measurement for both emission lines at the excitation power for which their intensity is roughly equal, under cw and pulsed excitation (line B). All show clear antibunching. This proves that increased extraction efficiency of $(13.3\pm2)$% and good single-photon purity can be combined within a single structure of a relatively simple design – QD embedded in a cylindrical mesa.

In conclusion, we demonstrated fabrication and characteristics of semiconductor quantum emitters which combine features important in view of applications in quantum communication in the fiber networks, namely: i) photon emission rate in the range of 5 MHz (single QD transition), ii) triggered single-photon emission under non-resonant excitation, iii) emission in the telecom C band spectral range and iv) compatibility with semiconductor technology due to epitaxial and embedded character of investigated InAs/InP nanostructures. Abovementioned emission rate is achieved thanks to increased extraction efficiency which was demonstrated to be on the level of (13.3±2)% into 0.4 numerical aperture at approx. 1560 nm emission wavelength close to the center of the telecom C-band. This is provided by combination of the bottom InP/InGaAlAs DBR section and cylindrical photonic confinement structure realized non-deterministically. Therefore, the proposed approach will benefit strongly from application of a deterministic technology platform [Rodt et al. (2020)].

**Acknowledgements**

This research was funded by the Foundation for Polish Science co-financed by the EU under the ERDF by project entitled „Quantum dot-based indistinguishable and entangled photon sources at telecom wavelengths" carried out within the HOMING programme. This work was also financially supported by the BMBF Project (Q.Link.X) and DFG (DeLiCom). We also acknowledge Andrei Kors for his assistance in the MBE growth process, Kerstin Fuchs and Dirk Albert for technical support.

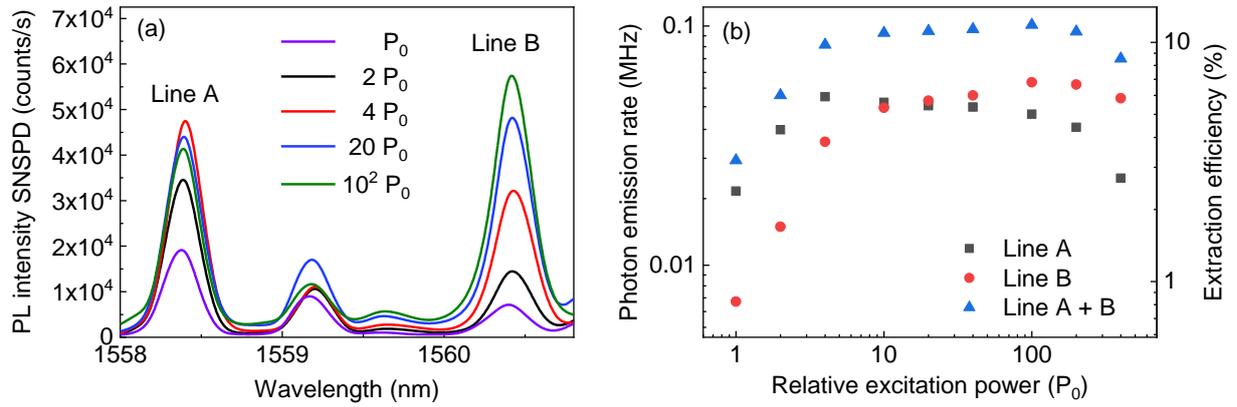

**FIG. 1** (a) Series of selected emission spectra for various excitation powers ($P_0 = 50$ nW) measured under pulsed excitation. (b) Photon emission rate (left axis) as a function of relative excitation power and extraction efficiency (right axis) for line A (black squares) and B (red dots) and sum of their intensities (blue triangles).

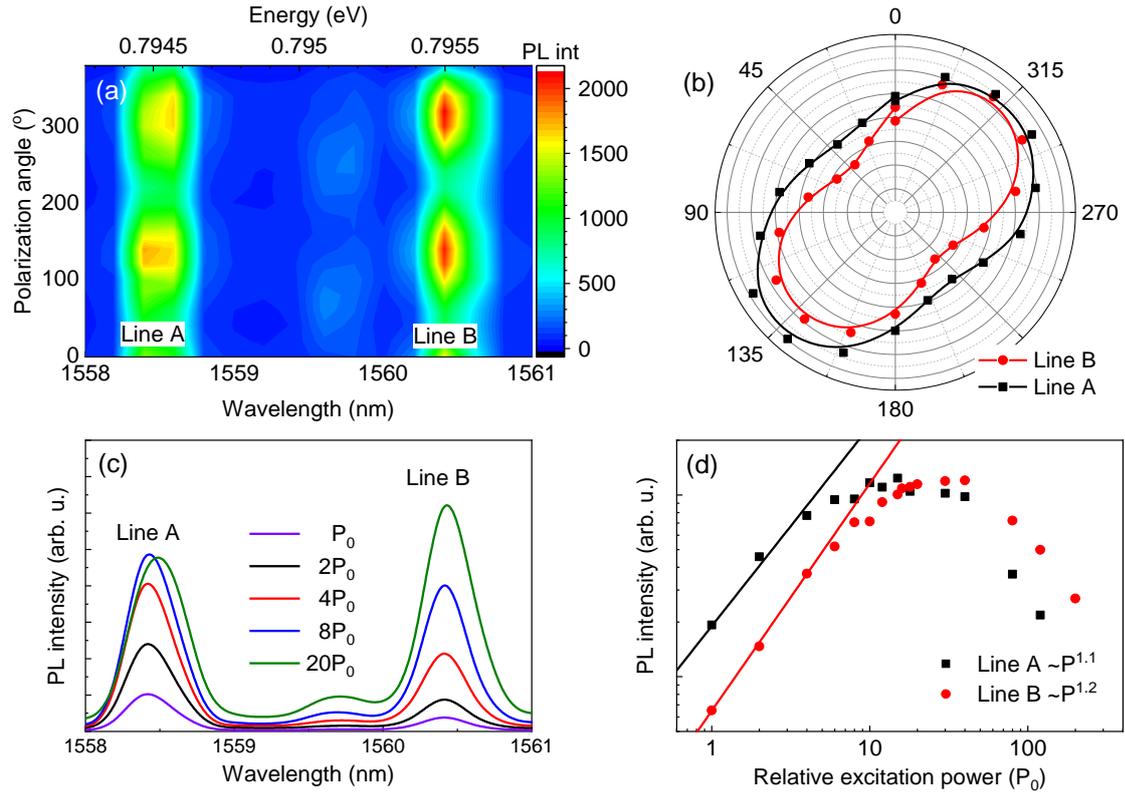

**FIG. 2** (a) Linear polarization map with color-coded PL intensity. (b) Polar plot of the PL intensity as a function of the linear polarization direction of the polarization analyzer for line A (black squares) and line B (red dots), solid lines - guide to the eye. (c) Series of selected emission spectra for various excitation powers ($P_0 = 50$ nW) measured under cw excitation. (d) Dependence of the integrated intensity from Gaussian fit to the emission lines on the relative excitation power for line A (black squares) and B (red dots) together with power function fit to the low excitation data (solid lines).

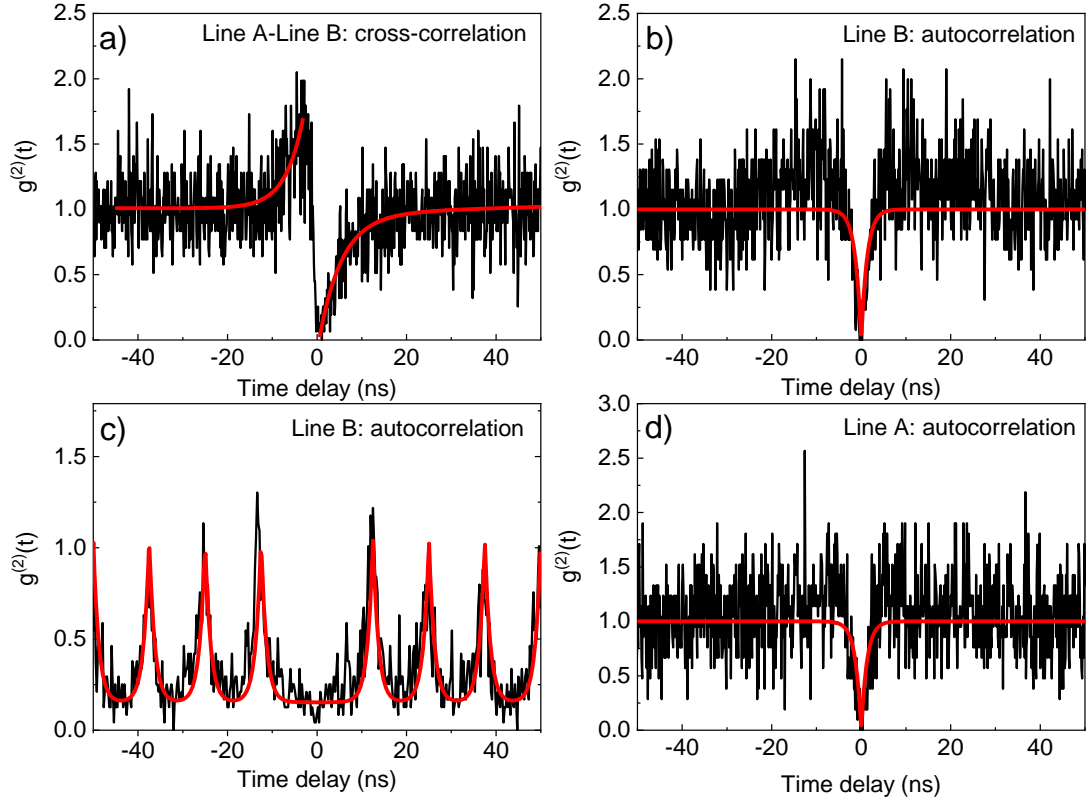

**FIG. 3** Second-order correlation function for non-resonant excitation (a) cross-correlation of line A and B under cw excitation, (b) and (c) autocorrelation of line B under cw and pulsed excitation, respectively. (d) Autocorrelation of line A for cw excitation. All histograms measured for excitation power at which the emission intensity of line B equals to emission intensity of line A. Red solid lines are fit to experimental data.